\documentclass[9pt,twocolumn,twoside]{opticajnl}
\journal{opticajournal} 

\usepackage{pdfpages}
\usepackage{lineno}

\title{Broadly tunable quantum-enhanced Raman microscopy for advancing bioimaging}

\author[1,2,*]{Dmitrii Akatev}
\author[3]{Yijian Meng}
\author[4]{Jonathan Brewer}
\author[2,5]{Maria Chekhova}
\author[1]{Ulrik L. Andersen}
\author[3,*]{Mikael Lassen}

\affil[1]{Center for Macroscopic Quantum States (bigQ), Department of Physics, Technical University of Denmark, Fysikvej 307, DK-2800 Kgs. Lyngby, Denmark}
\affil[2]{Max Planck Institute for the Science of Light, Staudtstr. 2, 91058 Erlangen, Germany}
\affil[3]{Danish Fundamental Metrology, Kogle Alle 5, DK-2970 Hørsholm, Denmark}
\affil[4]{University of Southern Denmark, Department of Molecular Biology and Biochemistry, Odense, Denmark}
\affil[5]{Friedrich-Alexander Universität Erlangen-Nürnberg,  91058 Erlangen, Germany}

\affil[*]{dmitrii.akatev@mpl.mpg.de or ml@dfm.dk}

\begin{abstract}
Stimulated Raman scattering (SRS) microscopy has emerged as a powerful technique for probing the spatiotemporal dynamics of molecular bonds with exceptional sensitivity, resolution, and speed. However, classically, its performance remains fundamentally constrained by optical shot noise, which imposes a strict limit on detection sensitivity and speed. Here, we demonstrate a quantum-enhanced SRS microscopy platform that circumvents this barrier by harnessing amplitude-squeezed light. Specifically, we generate a Stokes beam with $5.2~\mathrm{dB}$ of amplitude squeezing using traveling-wave optical parametric amplification in second-order nonlinear waveguides, and combine it with a tunable coherent pump to access vibrational modes spanning from $1000$ to $3100~\mathrm{cm}^{-1}$. Applied to quantum imaging of metabolites in biological tissue (pork muscle), our quantum-enhanced Raman microscope achieves an average noise suppression of $3.6~\mathrm{dB}$ and a $51\%$ enhancement in signal-to-noise ratio (SNR) -- to the best of our knowledge, the largest improvement reported to date in quantum-enhanced SRS microscopy of biological samples.
\end{abstract}

\begin{document}

\maketitle


\section{Introduction}

Quantum-enhanced optical sensing exploits nonclassical states of light to overcome the fundamental shot-noise limit, thereby enabling measurement sensitivities that surpass the capabilities of conventional optical methodologies \cite{Samantaray2017_SSNmicroscope, Ono2013_EntanglementEnhancedMicroscope}. Among nonclassical states of light, squeezed states constitute a particularly powerful resource, as they can redistribute quantum fluctuations to suppress noise in the quadrature of interest, thereby enhancing the signal-to-noise ratio (SNR) in precision measurements \cite{andersen201630,vahlbruch2008observation}. This capability allows the detection of extremely weak signals that would otherwise be obscured by classical noise limits. The potential of optical quantum-enhanced sensing has been demonstrated in several landmark experiments, most prominently through the deployment of squeezed light in gravitational-wave observatories \cite{LIGO-CITE,ganapathy2023broadband}. In microscopy and spectroscopy, such quantum-enhanced approaches open new avenues for probing molecular information at lower analyte concentrations, under reduced illumination intensities, or at accelerated acquisition  \cite{polzik1992spectroscopy,heng2025quantum,Lawrie:2019}.

Raman-based microscopy provides a powerful approach for probing molecular vibrations with high chemical specificity in a label-free and non-destructive manner \cite{Cheng2015}. However, the intrinsically weak nature of spontaneous Raman scattering limits both its sensitivity and acquisition speed, particularly in live-cell and clinical imaging contexts. To overcome these challenges, stimulated Raman scattering (SRS) microscopy was developed as a technique that resonantly drives molecular vibrations using synchronized pump and Stokes beams whose frequency difference is tuned to a specific vibrational mode of interest. By amplifying Raman signals by several orders of magnitude while preserving a linear dependence on molecular concentration, SRS enables quantitative, high-speed, and background-free chemical imaging \cite{Ozeki2012,Freudiger2008}. These capabilities have established SRS as a robust technique for spatially and chemically resolved imaging of lipids, proteins, metabolites, and pharmaceutical compounds in living systems, as well as for real-time biomedical applications, including intraoperative diagnostics \cite{hu2019biological,tipping2016stimulated}. Extending the principles of quantum-enhanced sensing to Raman-based microscopy provides a direct means to overcome the shot-noise limit that currently constrains SRS sensitivity. In conventional implementations, detection can be improved by increasing optical power or extending integration time, but for fragile biological tissues, excitation must remain below photodamage thresholds to preserve their native dynamics \cite{Zhang2022Phototoxic}. By employing squeezed probe states with noise levels below the standard shot-noise limit, quantum-enhanced Raman microscopy achieves enhanced sensitivity and faster acquisition rates, thereby minimizing photodamage in biological samples~\cite{Terrasson:24}. Beyond Raman-based approaches, close to 4~dB of quantum enhancement has been demonstrated in continuous-wave (CW) Brillouin microscopy, derived from small regions within the scanned area, where two-mode intensity-difference squeezed light generated via four-wave mixing in a Rb vapor cell was used to probe the mechanical properties of biological samples~\cite{li2024harnessing}. While Raman spectroscopy provides chemically specific biochemical contrast and is generally more suitable for bioimaging, Brillouin microscopy delivers complementary information on the viscoelastic and mechanical properties of the sample.

One of the first experimental proof-of-concept demonstrations of quantum-enhanced SRS (QE-SRS) employed CW amplitude-squeezed light, achieving a noise reduction of 3.6~dB at 2950~cm$^{-1}$ in polymer samples \cite{deAndrade:20}. Subsequently, pulsed amplitude squeezing enabled sub-shot-noise imaging of molecular bonds within live yeast cells, yielding a 1.3~dB noise reduction at 2850~cm$^{-1}$ \cite{Casacio2021}. Additionally, a quantum-enhanced balanced detection (QE-BD) scheme extended operation into the high-power regime (>30~mW), demonstrating a 3.11~dB noise reduction compared to classical detection strategies \cite{HP_balanced_Ozeki}. Beyond sensitivity improvements, squeezed light has also been leveraged to enhance imaging capabilities \cite{Gong:23}, while integration of bright squeezed probe beams into live-cell SRS microscopy enabled 20\% faster imaging with reduced photodamage \cite{Terrasson:24}. All of these studies demonstrate remarkable improvements offered by QE-SRS compared to conventional implementations. However, previous studies have been limited to a narrow subset of Raman vibrational modes. In Refs.~\cite{Terrasson:24,Casacio2021}, measurements were confined to the CH-stretch region, where overlapping Raman peaks from multiple biomolecules (e.g., lipids and fats) reduce chemical specificity. Similarly, Ref.~\cite{deAndrade:20} investigated simple polymer samples using a continuous-wave (CW) SRS configuration, which further restricts the extraction of detailed molecular information. Within the CH-stretch region, the Raman signatures of lipids, proteins, and nucleic acids exhibit substantial spectral overlap, making it difficult to discriminate between them. By extending measurements into the fingerprint region (500–1800~cm$^{-1}$), these overlapping contributions can be resolved, allowing unambiguous identification of distinct biomolecular vibrations and enabling a far richer and more specific chemical characterization of biological tissues.

In this work, we present a quantum-enhanced SRS microscope that employs a bright amplitude-squeezed Stokes beam as the probe field. The squeezed beam provides more than 5 dB of noise reduction while maintaining an optical power of ~3.75 mW, making it compatible with picosecond SRS imaging. Integrated into a tunable SRS platform, it enables broadband operation across both the fingerprint region (1450–1650~cm$^{-1}$) and the CH-stretch region (2800–3100~cm$^{-1}$), allowing chemically specific mapping of polystyrene reference samples and biological tissue with enhanced, achieving an approximately three-orders-of-magnitude improvement in material sensitivity compared to the CW approach reported in Ref.~\cite{deAndrade:20}. Whereas previous quantum-enhanced SRS demonstrations were confined to the CH-stretch region, the present work achieves, for the first time, quantum-enhanced vibrational imaging spanning both spectral regions, thereby enabling chemically distinguishable imaging in biologically relevant environments.

\section{Stimulated Raman microscopy setup}
\subsection{Full scheme}

\begin{figure}[htbp]
    \centering
    \includegraphics[width=\linewidth]{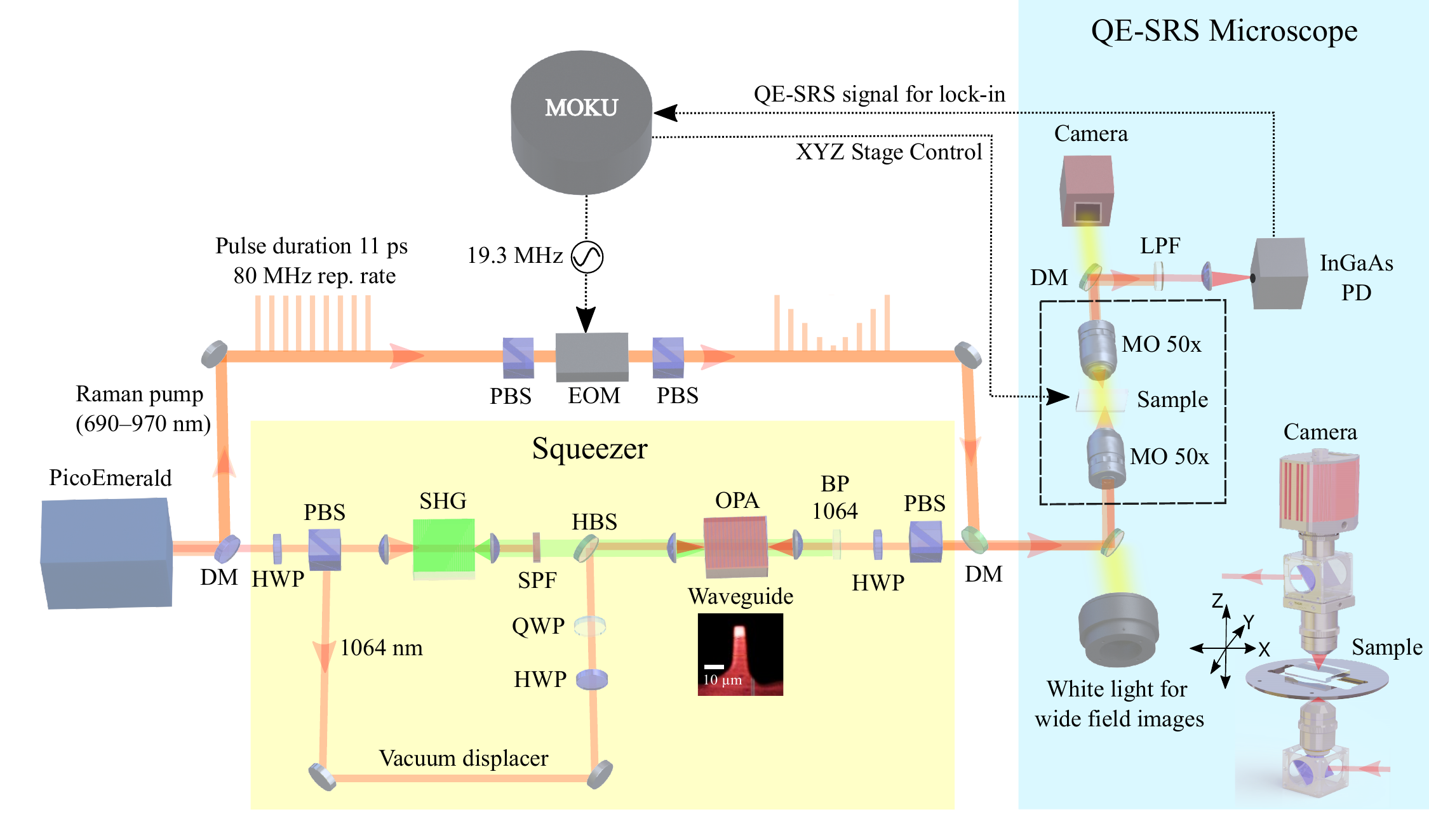}
    \caption{Block diagram of the main parts of the quantum-enhanced SRS microscopy setup. PicoEmerald: 
    pulsed laser (11 ps@80Mhz rep. rate). DM: dichroic mirrors. OPA: Squeezer, source of amplitude-squeezed light. SHG: second harmonic generation source for pumping OPA. PBS: polarizing beam splitter. HWP: Half-wave plates. EOM: electro-optic modulator for generating a 19.3 MHz modulation on the Raman pump. MO 50$\times$: microscope objective 50$\times$. BP 1064: 1064~nm bandpass filter. HBS: harmonic beam splitter. LPF: long-pass optical filter. SPF: short-pass optical filter. XYZ stage: 3D raster-scanning stage. Detector: InGaAs PD, home-made resonant photodetector. Moku: field-programmable gate array for system control and data acquisition. }
    \label{fig:main_setup}
\end{figure}

Fig.~\ref{fig:main_setup} shows the block diagram of the main parts of our quantum-enhanced stimulated Raman scattering (QE-SRS) microscope. The primary light source is a picoEmerald laser (APE Berlin), which delivers picosecond pulsed light suitable for accessing a wide range of Raman vibrational modes. It provides a broadly tunable Raman pump beam (690–970 nm) with up to 1000~mW of average optical power in the whole range, generated by an internal optical parametric oscillator (OPO), and a fixed-wavelength Stokes (coherent) beam at 1064 nm. Both beams have a pulse duration of 11 ps (full width at half maximum) and 80~MHz repetition rate. The Raman pump beam is separated from the 1064 nm Stokes beam by a dichroic mirror (DM, Thorlabs DMLP1000). Subsequently, amplitude modulation at $19.3~\mathrm{MHz}$ is applied to the Raman pump beam using a combination of polarizing beam splitters (PBS) and an electro-optical modulator (EOM, Thorlabs EO-AM-R-20-C1). The 1064 nm Stokes beam is directed to the squeezer to generate amplitude-squeezed light. The squeezer module enables switching between classical and squeezed-light configurations for the Stokes beam. The modulated Raman pump is then recombined with the amplitude-squeezed Stokes beam on another dichroic mirror DM. After recombination, both beams are directed into a custom-built microscope system. The microscope includes two 50$\times$ microscope objectives (MO 50$\times$, Thorlabs LMH-50X-1064), which focus the Stokes beam to a spot size of approximately 1.6~$\mu$m. Each objective lens exhibits approximately 98~\% transmission at 1064~nm. For precise and reliable sample positioning, we employ a custom-designed 3D raster-scanning stage (based on 8MTF-200-B43-MEN4-LEN1-025, Standa). An integrated white-light source and a wide-field imaging camera provide straightforward visualization and alignment of the sample prior to quantum-enhanced SRS measurements.

Following interaction with the sample, the Raman pump and Stokes beams are collected by the second MO 50$\times$. The Stokes beam is then separated from the Raman pump using a combination of a dichroic mirror (DM, Thorlabs DMSP1020B) and a long-pass filter (LPF, Thorlabs FELH1000) and directed onto a custom-built photodetector (InGaAs PD), adapted from the design of Ref.~\cite{Casacio2021}. The detector uses an InGaAs PIN Hamamatsu G10899-003K photodiode with a measured quantum efficiency of 97~\%. The detector is resonantly tuned to amplify signals at the 19.3~MHz modulation frequency. The SRS signal is acquired using a digital lock-in amplifier implemented on the Moku:Lab platform (MOKU, Liquid Instruments), with a 12-bit vertical resolution, and detection bandwidth of 2 kHz around the demodulation frequency achieving low-noise signal recovery. In addition, the Moku:Lab is used to generate the 19.3~MHz drive signal for the EOM, which enables precise amplitude modulation of the pump beam for SRS detection. For synchronization between the XYZ scanning stage and the data acquisition, we implement a custom Python-based control script, achieving a scan rate of 15 data points per second. While sufficient for proof-of-concept measurements, this rate is modest compared to commercial high-speed SRS systems. The Moku acquisition window, consisting of 1024 points, corresponds to approximately $\sim$10~ms. Therefore, the limitation arises primarily from the need to synchronize data acquisition with stage positioning, with the dwell time per pixel ($\sim$66~ms) being largely dominated by communication and data handling overhead in the Python-based acquisition framework.

\subsection{Preparation of amplitude-squeezed light}

To obtain an amplitude-squeezed Stokes beam, we employed a single-pass optical parametric amplifier (OPA) based on a periodically poled lithium niobate (PPLN) waveguide \cite{Eto:11,Amari:23}. A simplified schematic of the source is shown in Fig.~\ref{fig:main_setup} (the yellow box). The squeezed-vacuum (SV) state was generated by pumping the PPLN waveguide with a 532~nm beam. To realize an amplitude-squeezed state, the SV was displaced by a coherent beam (displacing beam) using an asymmetric beam splitter composed of a half-wave plate (HWP) and a polarizing beam splitter (PBS). For stable displacement, the relative phase between the SV and the displacing beam was actively stabilized using an auxiliary reference beam introduced into the squeezer module, which provided the error signal for a feedback loop driving active phase control.
As a result, we achieved a 5.2~$\pm$~0.2~dB reduction of amplitude quadrature noise with a displacement of 3.75~mW using an OPA pump power of 2.5~mW. A detailed description of the optical layout, phase-stabilization scheme, and measurement procedure is provided in Supplementary~1, Section~1.

We validated the noise-reduction performance of the prepared amplitude-squeezed state by an SRS measurement. Using a simplified experimental setup, we recorded the SRS signal from a 1~mm-thick polystyrene layer at 3050~cm$^{-1}$, corresponding to the vibrational modes of the C--H bonds within aromatic rings and the associated ring structures. This measurement yielded an improvement of approximately 4~dB in the background noise level, representing one of the highest reported measured enhancements for QE-SRS on polystyrene~\cite{deAndrade:20}. Further experimental details are provided in Supplementary~1, Section~2.

\subsection{Tunability of QE-SRS microscope}

\begin{figure}[htbp]
    \centering
    \includegraphics[width=\linewidth]{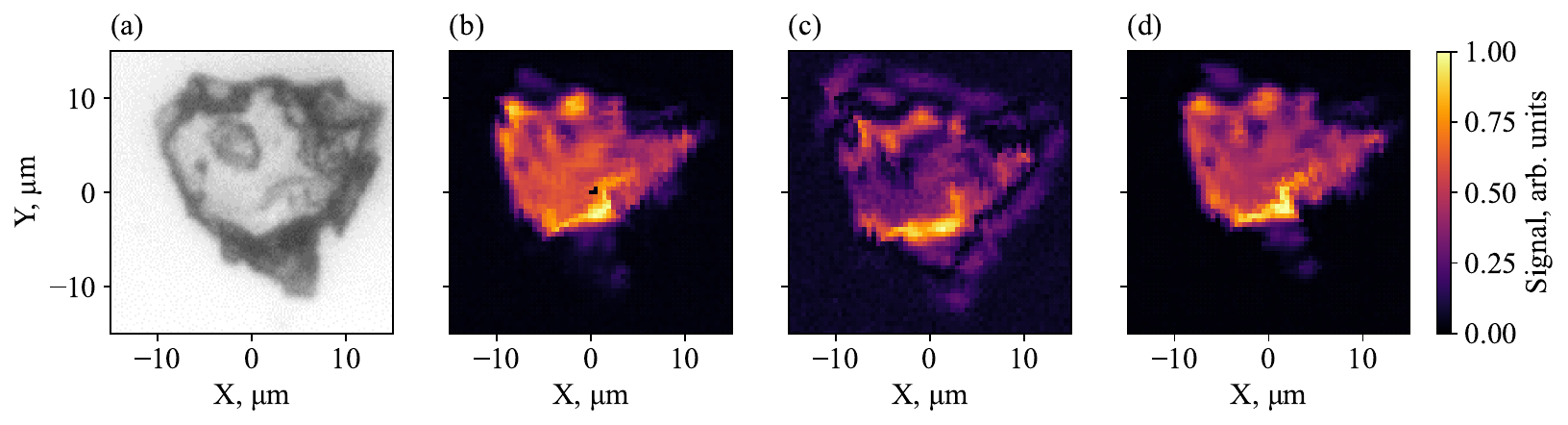}
    \caption{(a) White-light wide-field image of a polystyrene microparticle; 
    SRS images of the same particle at Raman shifts
    (b) 1003~cm$^{-1}$ (symmetric ring-breathing mode); 
    (c) 1600~cm$^{-1}$ (skeletal C=C stretching mode); 
    (d) 3050~cm$^{-1}$ (aromatic C–H stretching mode).}
    \label{fig:Common_plot_pol_bead}
\end{figure}

In addition to the high detected squeezing level, our system provides broad spectral tunability, allowing access to both the fingerprint and functional group regions. 
This broad tunability is critical for biochemical fingerprinting because the fingerprint and CH-stretch regions contain distinct vibrational markers associated with different molecular species. As an example of the tunability, Fig.~\ref{fig:Common_plot_pol_bead} presents SRS imaging of a polystyrene microparticle obtained using a coherent (classical) Stokes beam. Polystyrene was selected as a well-established reference material owing to its strong Raman signals at well defined Raman shifts across a broad spectral range~\cite{Kerdoncuff:19}.

We investigated three of the most prominent Raman lines of polystyrene: 3050~cm$^{-1}$, corresponding to the aromatic C–H stretching mode; 1600~cm$^{-1}$, associated with the skeletal C=C stretching mode of the phenyl ring; and 1003~cm$^{-1}$, corresponding to the symmetric ring-breathing mode of the phenyl group. To match the targeted Raman lines, the pump wavelength from the picoEmerald laser was tuned from 803.5~nm to 961~nm, covering the spectral range from the C–H stretching mode to the ring-breathing mode. The Raman pump power was set to approximately 34 mW, while the coherent Stokes beam power was maintained at around 3.75 mW to ensure a sufficient signal level. The SRS images obtained with our home-built microscopy system show good agreement with the corresponding white-light wide-field image, confirming both the spatial resolution and spectral tunability of the system.

\section{Quantum-enhanced SRS bioimaging} 

\subsection{QE-SRS on a pork muscle tissue}

To demonstrate the applicability of our QE-SRS platform for biological imaging, we investigated a pork muscle tissue sample. The tissue was sectioned into 10~$\mu$m-thick slices and mounted on a cover glass. The sample was sealed with water using an additional cover glass, forming a cuvette-like structure that preserved hydration during imaging. This configuration also facilitated heat dissipation induced by optical illumination and improved the structural integrity of the prepared sample. Moreover, the gradual transition in refractive index (glass $\rightarrow$ sample $\rightarrow$ water $\rightarrow$ glass) helped suppress optical scattering typically arising from refractive index mismatches in biological tissues.

\begin{figure}
    \centering
    \includegraphics[width=\linewidth]{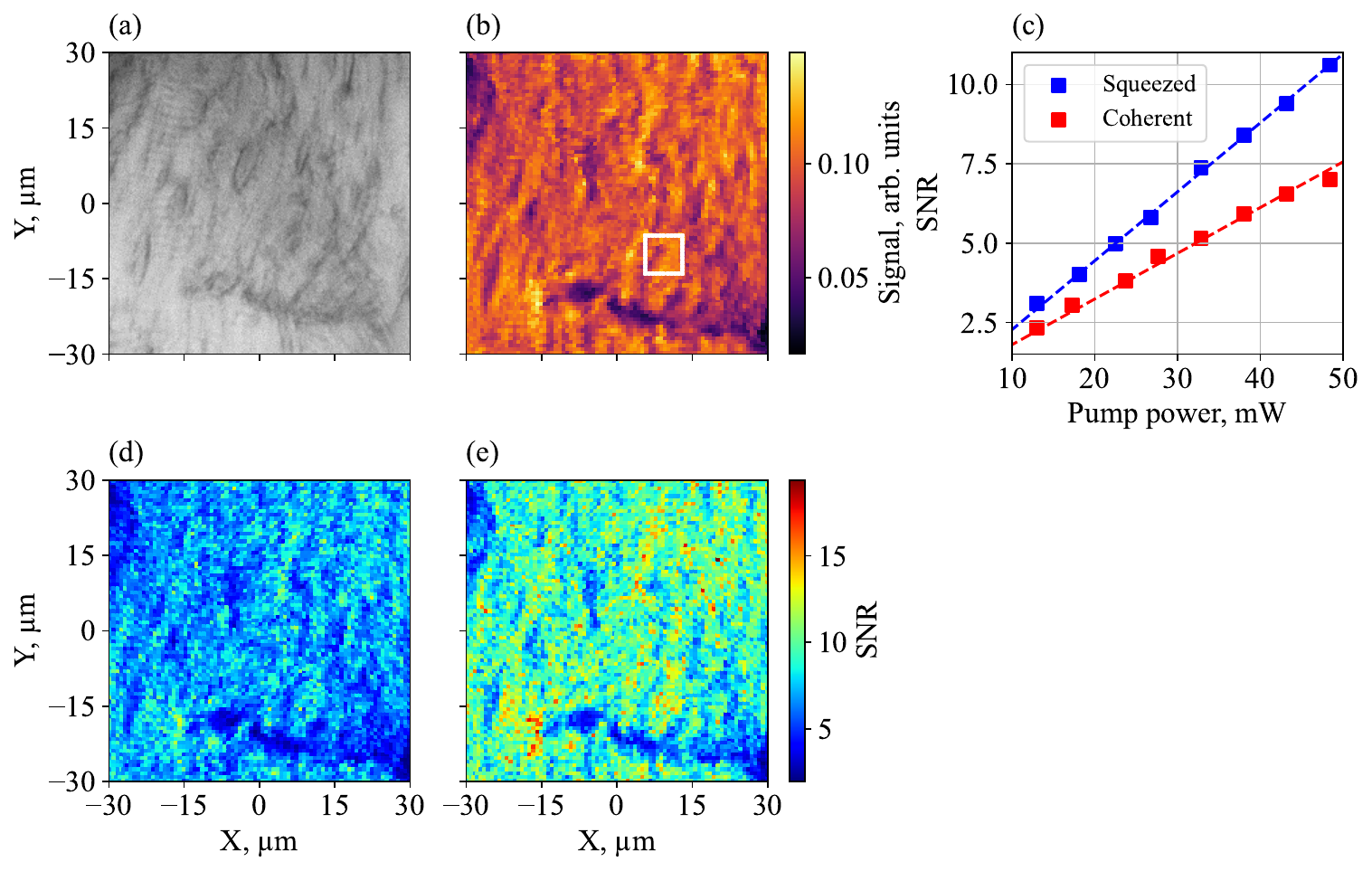}
    \caption{(a) White-light wide-field image of the pork muscle tissue sample. (b) SRS image acquired at 2940~cm$^{-1}$ in the same region as the wide-field image. (c) Signal-to-noise ratio of the SRS signal from the region indicated by the white square in panel (b), plotted as a function of Raman pump power for squeezed (blue) and coherent (red) Stokes beams of equal power. Maps of SRS signal-to-noise ratio for (d) coherent and (e) squeezed Stokes beams.
}
    \label{fig:for_power_dep_full_view}
\end{figure}

The Raman band at 2940~cm$^{-1}$ originates from CH$_2$/CH$_3$ stretching modes present in both proteins and lipids and therefore serves as an indicator of overall protein–lipid content in tissue. 
The measurement was performed using an average optical power of 32~mW for the Raman pump and 3.75~mW for the Stokes beam focused on the sample. The white-light wide-field image is shown in Fig.~\ref{fig:for_power_dep_full_view}(a). Spatial mapping of the muscle structure was achieved via raster scanning with a step size of 750~nm. The measured SRS signal (Fig.~\ref{fig:for_power_dep_full_view}(b)) clearly reproduces the structural features of the muscle tissue. Fig.~\ref{fig:for_power_dep_full_view}(d),(e) further illustrate the enhancement in SNR achieved by employing amplitude-squeezed light. In the case of SRS gain detection, the SNR is ~\cite{deAndrade:20}
{\begin{equation}
\text{SNR} = \frac{I_\text{SRS}}{\sqrt{\Delta I^2_\text{s}}},
\end{equation}
where $I_\text{SRS}\sim I_\text{p} I_\text{s}$ is the SRS signal intensity, $I_\text{p}$ and $I_\text{s}$ are pump and Stokes beam intensities, respectively, and $\Delta I^2_\text{s}$ is the variance of the detected Stokes beam intensity that quantifies the noise level. By using amplitude-squeezed light, we are able to surpass the shot-noise level, thereby enhancing the SNR in proportion to the inverse square root of the noise reduction factor. 
In Fig.~3(d) and (e), we observed an SNR improvement across the entire investigated area ($60\times60~\mu\mathrm{m}^2$),
\begin{equation}
    \delta = \frac{\mathrm{SNR}_{sqz}}{\mathrm{SNR}_{coh}}\approx 1.46,
\end{equation}
which corresponds to an approximately $46\%$ increase in SNR. Using this value, we can estimate the corresponding noise reduction as
\begin{equation}
    \text{S} = -20\log_{10}\left(\frac{1}{\delta}\right)
    = -20\log_{10}\left(\frac{1}{1.46}\right)\approx 3.3~\mathrm{dB},
\end{equation}
which indicates a noise reduction of $3.3~\mathrm{dB}$ compared to the coherent Stokes beam configuration.
The initial $5.2~\mathrm{dB}$ of squeezing was reduced to $3.3~\mathrm{dB}$ at the detector due to 24\% optical loss arising from optics and sample transmission. 

In addition, we investigated the scaling behavior of the SNR with respect to the Raman pump power. As discussed above, the SNR is expected to scale linearly with pump power. To validate this, we measured the SNR of the SRS signal within the white square region shown in Fig.~\ref{fig:for_power_dep_full_view}(b). The results are presented in Fig.~\ref{fig:for_power_dep_full_view}(c). Here, the SNR was averaged over a $7 \times 7~\mu$m$^2$ area, while the pump power was varied from 12~mW to 49~mW with the Stokes beam power fixed at 3.75~mW. The blue and red dashed lines correspond to linear fits of the acquired data for coherent (red trace) and amplitude-squeezed (blue trace) Stokes beams, respectively. The increased slope observed in the squeezed-light case corresponds to an estimated noise reduction of approximately $S\approx3.3~\mathrm{dB}$, yielding a 46\% ($\delta \approx 1.46$) improvement in SNR, which is consistent with the enhancements observed in panels~(d) and~(e).

\subsection{Multifrequency biological imaging}

\begin{figure}[htbp]
    \centering
    \includegraphics[width=\linewidth]{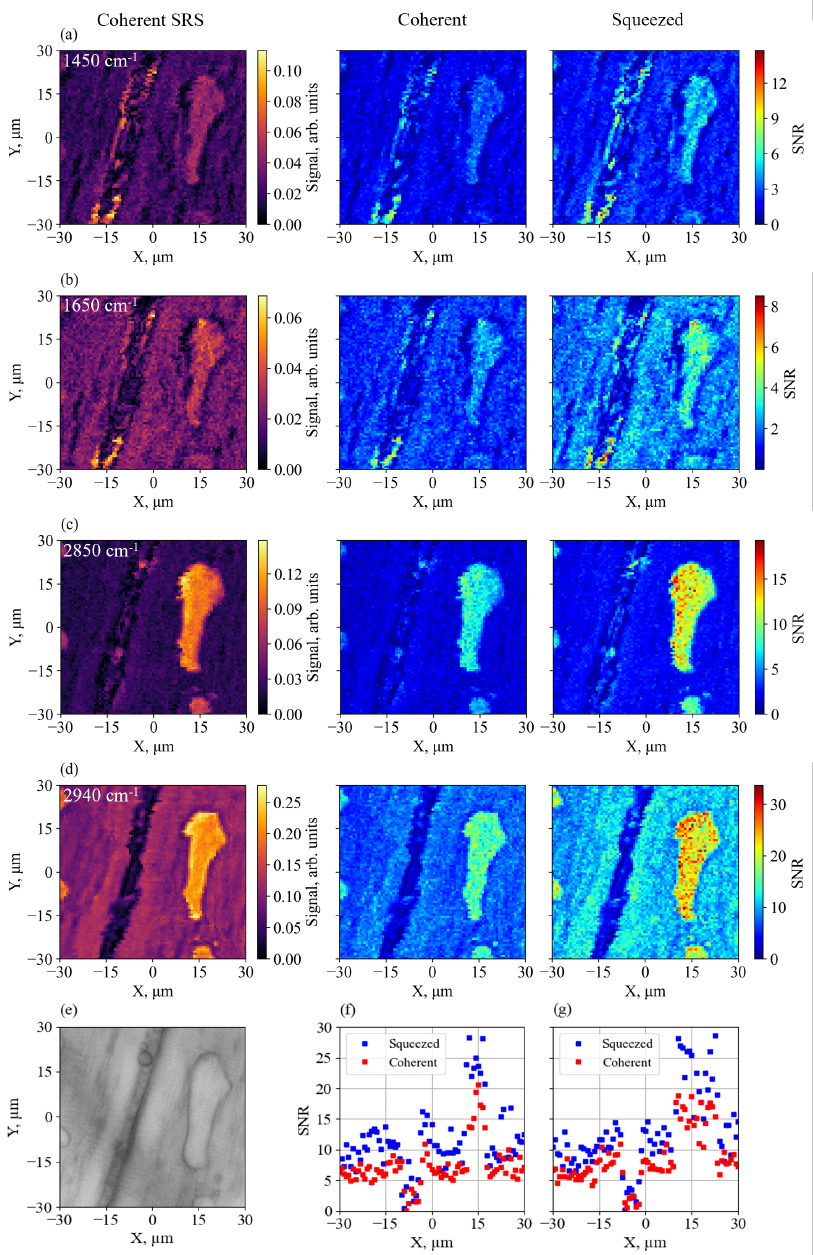}
    \caption{
        Quantum-enhanced SRS images of pork muscle tissue acquired at:
        (a) 1450~cm$^{-1}$, 
        (b) 1650~cm$^{-1}$, 
        (c) 2850~cm$^{-1}$, and 
        (d) 2940~cm$^{-1}$. 
        Left panels show the SRS intensity obtained using a coherent Stokes beam; middle and right panels display the corresponding SNR measurements for coherent and amplitude-squeezed Stokes beams, respectively. 
        (e) White-light wide-field image of the sample. 
        (f) and (g) Cross-sectional profiles of the SNR along Y = 0 $\mu m$ and Y = 15 $\mu m$ from panel (d), respectively, comparing the coherent (red) and amplitude-squeezed (blue) Stokes beam configurations.
    }
    \label{fig:muscle_SRS}
\end{figure}

To demonstrate chemical specificity, we performed SRS imaging at four vibrational bands characteristic of proteins and lipids.
Explicitly, we investigated several prominent Raman vibrational modes: $2940~\mathrm{cm^{-1}}$ (general C--H stretching in proteins, lipids, and nucleic acids), widely used as a marker of total protein and lipid content; $2850~\mathrm{cm^{-1}}$ (symmetric CH$_2$ stretching in lipids); $1650~\mathrm{cm^{-1}}$ (amide I band, primarily C=O stretching in proteins), a key indicator of protein content and secondary structure; and $1450~\mathrm{cm^{-1}}$ (CH$_2$/CH$_3$ bending in lipids and proteins), often used as a marker for lipid–protein balance. Together, these four bands give a complementary biochemical fingerprint of muscle tissue. For these measurements, the Raman pump was tuned to $810.5~\mathrm{nm}$, $816.5~\mathrm{nm}$, $905~\mathrm{nm}$, and $921.8~\mathrm{nm}$ with corresponding powers at the sample of 32~mW, 32~mW, 30~mW, and 38~mW, respectively. The Stokes beam power was maintained at $\sim 3.75~\mathrm{mW}$ in all cases. However, systematically higher SNR were observed at the CH-stretch bands (2850 and 2940~cm$^{-1}$) compared to the fingerprint modes. This enhancement is attributed to the larger Raman cross-sections and the higher abundance of CH groups in tissue lipids and proteins, particularly in lipid-rich structures such as cell membranes and adipose tissue. In contrast, Fig.~\ref{fig:muscle_SRS} shows regions with SRS signals at 1450~cm$^{-1}$ and 1650~cm$^{-1}$, where the CH-stretch bands at 2850 and 2940~cm$^{-1}$ are absent. This spectral pattern indicates regions with elevated protein content and minimal lipid contribution, consistent with lean myofibrillar or connective tissue in muscle \cite{pezzotti2021raman,afseth2022raman}. The results presented in Fig.~\ref{fig:muscle_SRS} demonstrate a consistent enhancement in SNR across all examined Raman shifts when utilizing the amplitude-squeezed Stokes beam. A noise reduction of $3.6 \pm 0.14$~dB was achieved across all measured vibrational modes, corresponding to a $51 \pm 5\%$ improvement in SNR compared to the classical (coherent) Stokes configuration. Fig.~\ref{fig:muscle_SRS}(f) and (g) show cross-sections of the SNR maps. These cross-sections clearly illustrate the improvements achieved with squeezed light compared to coherent illumination.
Along with the observed SNR enhancement, the achieved squeezing of the Stokes beam directly improves the concentration sensitivity. In particular, since the concentration sensitivity scales with the noise amplitude \cite{deAndrade:20}, i.e. $\delta N \propto \sqrt{\Delta I_s^2}$, a noise suppression of $3.6~\mathrm{dB}$ corresponds to
\begin{equation}
\frac{\delta N_{sqz}}{\delta N_{coh}} = \sqrt{\frac{\Delta I_{sqz}^2}{\Delta I_{coh}^2}} \approx 0.66,
\end{equation}
which indicates that the minimum detectable concentration is reduced to $\approx 66\%$ of its coherent-state value, i.e., improved by approximately $34\%$.
This consistent enhancement underscores the robustness of quantum-assisted amplification in improving Raman signal quality. The observed gains highlight the potential of QE-SRS to extend the sensitivity of vibrational imaging in biological tissues, enabling more precise and reliable real-time biochemical analysis, even at lower optical powers or in samples with weak Raman signatures.

\section{Conclusion}

In summary, we have implemented quantum-enhanced broadband SRS microscopy for biological applications. To this end, we have first generated a pulsed squeezed vacuum beam using single-pass OPA in a PPLN waveguide. By combining the squeezed vacuum with a coherent displacement beam under carefully optimized spatial, temporal, and polarization overlap, we obtained a 5.2~dB amplitude-squeezed bright beam, providing a stable and controllable nonclassical light source for the use as the Stokes beam for QE-SRS microscopy. Together with the tunable Raman pump, the squeezed Stokes beam provided quantum-enhanced SRS microscopy. The setup demonstrated reliable operation with tunability across both the fingerprint (1450--1650~cm$^{-1}$) and functional group (2800--3100~cm$^{-1}$) regions, covering many of the most relevant molecular vibrational modes in biological systems.

We validated the performance of the quantum-enhanced SRS microscope by conducting measurements on both benchmark polystyrene samples and biological samples (pork muscle tissue). The biological measurements revealed an average noise reduction of 3.6~dB, corresponding to a 51\% improvement in SNR compared to classical detection. To the best of our knowledge, this represents the highest reported enhancement in QE-SRS microscopy applied to biological samples. 
Importantly, our results further support the applicability of squeezed-light injection in picosecond-pulsed SRS microscopy for complex biological tissue environments, while also indicating clear potential for further optimization.
Along with strong squeezing, we also report, for the first time, QE-SRS across both regions in the fingerprint and high-wavenumber CH-stretch regions. Our approach provides quantum SNR enhancement in both regions, enabling sensitive and label-free imaging and identification of important biomarkers, such as proteins, lipids, amid and nucleic acids in biological samples.

Looking forward, further improvements in coupling efficiency, waveguide engineering, and pump power stabilization are expected to increase the achievable squeezing levels and long-term stability of the source.Integration with advanced lock-in detection schemes, optimized data acquisition and synchronization, and fast-scanning microscopy incorporating galvanometric scanners or resonant scanners could enable real-time, quantum-enhanced imaging of living biological samples, thereby extending the approach to practical biomedical applications. Moreover, combining squeezed-light excitation with computational imaging and machine learning approaches may open new possibilities for super-resolved Raman microscopy with reduced photo damage and enhanced molecular contrast.

\section*{Acknowledgment} 
This research was funded by the Danish Agency for Institutions and Educational Grants and QuRaman project under QuantERA and Eureka project QMIC supported by Innovation Fund Denmark (1116-00003B and 4340-00008B), the German Federal Ministry of Education and Research under the grant number 13N16359, the Danish National Research Foundation (bigQ, DNRF142) and the Novo Nordisk Foundation (CBQS, NNF24SA0088433).

\section*{Disclosures}
The authors declare no conflicts of interest.

\section*{Data availability}
Data underlying the results presented in this paper are not publicly available but may be obtained from the authors upon reasonable request.

\section*{Supplemental document}
See Supplementary 1 for supporting content.

\bibliography{sample}

@article{Gong:23,
author = {Li Gong and Shulang Lin and Zhiwei Huang},
journal = {Opt. Lett.},
number = {24},
pages = {6516--6519},
publisher = {Optica Publishing Group},
title = {Super-resolution stimulated Raman scattering microscopy enhanced by quantum light and deconvolution},
volume = {48},
month = {Dec},
year = {2023},
url = {https://opg.optica.org/ol/abstract.cfm?URI=ol-48-24-6516},
doi = {10.1364/OL.509616},
}

@article{Terrasson:24,
author = {Alex Terrasson and Nicolas P. Mauranyapin and Catxere A. Casacio and Joel Q. Grim and Kai Barnscheidt and Boris Hage and Michael A. Taylor and W. P. Bowen},
journal = {Opt. Express},
number = {21},
pages = {36193--36206},
publisher = {Optica Publishing Group},
title = {Fast biological imaging with quantum-enhanced Raman microscopy},
volume = {32},
month = {Oct},
year = {2024},
url = {https://opg.optica.org/oe/abstract.cfm?URI=oe-32-21-36193},
doi = {10.1364/OE.523956},
}

@article{Lawrie:2019,
author = {Lawrie, B. J. and Lett, P. D. and Marino, A. M. and Pooser, R. C.},
title = {Quantum Sensing with Squeezed Light},
journal = {ACS Photonics},
volume = {6},
number = {6},
pages = {1307-1318},
year = {2019},
doi = {10.1021/acsphotonics.9b00250},

URL = { 
    
        https://doi.org/10.1021/acsphotonics.9b00250
    
    

},
eprint = { 
    
        https://doi.org/10.1021/acsphotonics.9b00250
}
}

@article{deAndrade:20,
author = {Rayssa B. de Andrade and Hugo Kerdoncuff and Kirstine Berg-S{\o}rensen and Tobias Gehring and Mikael Lassen and Ulrik L. Andersen},
journal = {Optica},
number = {5},
pages = {470--475},
publisher = {Optica Publishing Group},
title = {Quantum-enhanced continuous-wave stimulated Raman scattering spectroscopy},
volume = {7},
month = {May},
year = {2020},
url = {https://opg.optica.org/optica/abstract.cfm?URI=optica-7-5-470},
doi = {10.1364/OPTICA.386584},
}

@article{Casacio2021,
  author       = {Casacio, Catxere A. and Madsen, Lars S. and Terrasson, Alex and Waleed, Muhammad and Barnscheidt, Kai and Hage, Boris and Taylor, Michael A. and Bowen, Warwick P.},
  title        = {Quantum-enhanced nonlinear microscopy},
  journal      = {Nature},
  year         = {2021},
  volume       = {594},
  number       = {7862},
  pages        = {201--206},
  doi          = {10.1038/s41586-021-03528-w},
}

@article{Eto:11,
author = {Yujiro Eto and Akane Koshio and Akito Ohshiro and Junichi Sakurai and Keiko Horie and Takuya Hirano and Masahide Sasaki},
journal = {Opt. Lett.},
number = {23},
pages = {4653--4655},
publisher = {Optica Publishing Group},
title = {Efficient homodyne measurement of picosecond squeezed pulses with pulse shaping technique},
volume = {36},
month = {Dec},
year = {2011},
url = {https://opg.optica.org/ol/abstract.cfm?URI=ol-36-23-4653},
doi = {10.1364/OL.36.004653},
}

@article{Amari:23,
author = {Jorge Amari and Junnosuke Takai and Takuya Hirano},
journal = {Opt. Continuum},
number = {4},
pages = {933--941},
publisher = {Optica Publishing Group},
title = {Highly efficient measurement of optical quadrature squeezing using a spatial light modulator controlled by machine learning},
volume = {2},
month = {Apr},
year = {2023},
url = {https://opg.optica.org/optcon/abstract.cfm?URI=optcon-2-4-933},
doi = {10.1364/OPTCON.484295},
}

@article{LIGO-CITE,
  title        = {Enhanced sensitivity of the {LIGO} gravitational wave detector by using squeezed states of light},
  author       = {Aasi, J. and Abadie, J. and Abbott, B.~P. and Abbott, R.~S. and Abbott, T.~D. and Abernathy, M.~R. and Adams, C. and Adams, T. and Addesso, P. and Adhikari, R.~X. and Affeldt, C. and Aguiar, O.~D. and Ajith, P. and Allen, B. and Amador Cerón, E. and Amariutei, D. and Anderson, S.~B. and Anderson, W.~G. and Arai, K. and ... and Zweizig, J.~G.},
  journal      = {Nature Photonics},
  volume       = {7},
  number       = {8},
  pages        = {613--619},
  year         = {2013},
  doi          = {10.1038/nphoton.2013.177}
}

@article{ganapathy2023broadband,
  title={Broadband quantum enhancement of the {LIGO} detectors with frequency-dependent squeezing},
  author={Ganapathy, D and Jia, W and Nakano, M and Xu, V and Aritomi, N and Cullen, T and Kijbunchoo, N and Dwyer, SE and Mullavey, A and McCuller, L and others},
  journal={Physical Review X},
  volume={13},
  number={4},
  pages={041021},
  year={2023},
  publisher={APS}
}

@article{andersen201630,
  title={30 years of squeezed light generation},
  author={Andersen, Ulrik L and Gehring, Tobias and Marquardt, Christoph and Leuchs, Gerd},
  journal={Physica Scripta},
  volume={91},
  number={5},
  pages={053001},
  year={2016},
  publisher={IOP Publishing}
}

@article{vahlbruch2008observation,
  title={Observation of squeezed light with 10-dB quantum-noise reduction},
  author={Vahlbruch, Henning and Mehmet, Moritz and Chelkowski, Simon and Hage, Boris and Franzen, Alexander and Lastzka, Nico and Gossler, Stefan and Danzmann, Karsten and Schnabel, Roman},
  journal={Physical review letters},
  volume={100},
  number={3},
  pages={033602},
  year={2008},
  publisher={APS}
}

@article{Kerdoncuff:19,
author = {Hugo Kerdoncuff and Mikael Lassen and Jan C. Petersen},
journal = {Opt. Lett.},
number = {20},
pages = {5057--5060},
publisher = {Optica Publishing Group},
title = {Continuous-wave coherent Raman spectroscopy for improving the accuracy of Raman shifts},
volume = {44},
month = {Oct},
year = {2019},
url = {https://opg.optica.org/ol/abstract.cfm?URI=ol-44-20-5057},
doi = {10.1364/OL.44.005057},
}

@article{Samantaray2017_SSNmicroscope,
  author    = {Samantaray, N. and Ruo-Berchera, I. and Meda, A. and Genovese, M. and others},
  title     = {Realization of the first sub-shot-noise wide field microscope},
  journal   = {Light: Science \& Applications},
  year      = {2017},
  volume    = {6},
  pages     = {e17005},
  doi       = {10.1038/lsa.2017.5},
}

@article{Ono2013_EntanglementEnhancedMicroscope,
  author    = {Ono, Takafumi and Okamoto, Ryo and Takeuchi, Shigeki},
  title     = {An entanglement-enhanced microscope},
  journal   = {Nature Communications},
  year      = {2013},
  volume    = {4},
  pages     = {2426},
  doi       = {10.1038/ncomms3426},
}

@article{Ozeki2012,
  author    = {Ozeki, Yasuyuki and Umemura, Wataru and Otsuka, Yusuke and Satoh, Satoshi and Hashimoto, Hirohiko and Sumimura, Kazuki and Nishizawa, Norihiko and Fukui, Keisuke and Itoh, Kazuyoshi},
  title     = {High-speed molecular spectral imaging of tissue with stimulated Raman scattering},
  journal   = {Nat. Photonics},
  volume    = {6},
  pages     = {845--851},
  year      = {2012},
  doi       = {10.1038/nphoton.2012.263}
}

@article{Freudiger2008,
  author    = {Freudiger, Christian W. and Min, Wei and Saar, Brian G. and Lu, Sijia and Holtom, Gary R. and He, Chang and Tsai, Jason C. and Kang, Jeon Woong and Xie, X. Sunney},
  title     = {Label-free biomedical imaging with high sensitivity by stimulated Raman scattering microscopy},
  journal   = {Science},
  volume    = {322},
  number    = {5909},
  pages     = {1857--1861},
  year      = {2008},
  doi       = {10.1126/science.1165758}
}

@article{Cheng2015,
  author  = {Cheng, Ji-Xin and Xie, X. Sunney},
  title   = {Vibrational spectroscopic imaging of living systems: An emerging platform for biology and medicine},
  journal = {Science},
  volume  = {350},
  number  = {6264},
  pages   = {aaa8870},
  year    = {2015},
  doi     = {10.1126/science.aaa8870}
}

@article{hu2019biological,
  title={Biological imaging of chemical bonds by stimulated Raman scattering microscopy},
  author={Hu, Fanghao and Shi, Lixue and Min, Wei},
  journal={Nature methods},
  volume={16},
  number={9},
  pages={830--842},
  year={2019},
  publisher={Nature Publishing Group US New York}
}

@article{tipping2016stimulated,
  title={Stimulated Raman scattering microscopy: an emerging tool for drug discovery},
  author={Tipping, WJ and Lee, Martin and Serrels, A and Brunton, VG and Hulme, AN},
  journal={Chemical Society Reviews},
  volume={45},
  number={8},
  pages={2075--2089},
  year={2016},
  publisher={Royal Society of Chemistry}
}

@article{li2024harnessing,
  title={Harnessing quantum light for microscopic biomechanical imaging of cells and tissues},
  author={Li, Tian and Cheburkanov, Vsevolod and Yakovlev, Vladislav V and Agarwal, Girish S and Scully, Marlan O},
  journal={Proceedings of the National Academy of Sciences},
  volume={121},
  number={45},
  pages={e2413938121},
  year={2024},
  publisher={National Academy of Sciences}
}

@article{heng2025quantum,
  title={Quantum-Enhanced Sensing with Squeezed Light: From Fundamentals to Applications},
  author={Heng, Xing and Zhang, Lingchen and Yin, Qingyun and Liu, Wei and Tang, Lulu and Zhai, Yueyang and Wei, Kai},
  journal={Applied Sciences},
  volume={15},
  number={18},
  pages={10179},
  year={2025},
  publisher={MDPI}
}

@article{polzik1992spectroscopy,
  title={Spectroscopy with squeezed light},
  author={Polzik, ES and Carri, J and Kimble, HJ},
  journal={Physical review letters},
  volume={68},
  number={20},
  pages={3020},
  year={1992},
  publisher={APS}
}

@article{pezzotti2021raman,
  title={Raman spectroscopy in cell biology and microbiology},
  author={Pezzotti, Giuseppe},
  journal={Journal of Raman Spectroscopy},
  volume={52},
  number={12},
  pages={2348--2443},
  year={2021},
  publisher={Wiley Online Library}
}

@article{afseth2022raman,
  title={Raman and near infrared spectroscopy for quantification of fatty acids in muscle tissue—a salmon case study},
  author={Afseth, Nils Kristian and Dankel, Katinka and Andersen, Petter Vejle and Difford, Gareth Frank and Horn, Siri Storteig and Sonesson, Anna and Hillestad, Borghild and Wold, Jens Petter and Tengstrand, Erik},
  journal={Foods},
  volume={11},
  number={7},
  pages={962},
  year={2022},
  publisher={MDPI}
}

@article{HP_balanced_Ozeki,
    author = {Xu, Zicong and Nitanai, Sho and Oguchi, Kenichi and Ozeki, Yasuyuki},
    title = {Pushing the sensitivity of stimulated Raman scattering microscopy with quantum light: Current status and future challenges},
    journal = {Applied Physics Letters},
    volume = {127},
    number = {4},
    pages = {040501},
    year = {2025},
    month = {07},
    issn = {0003-6951},
    doi = {10.1063/5.0273335},
    url = {https://doi.org/10.1063/5.0273335},
    eprint = {https://pubs.aip.org/aip/apl/article-pdf/doi/10.1063/5.0273335/20616741/040501_1_5.0273335.pdf},
}

@article{Zhang2022Phototoxic,
  author  = {Zhang, X. and Dorlhiac, G. and Landry, M. P. and Streets, A.},
  title   = {Phototoxic effects of nonlinear optical microscopy on cell cycle, oxidative states, and gene expression},
  journal = {Scientific Reports},
  volume  = {12},
  pages   = {18796},
  year    = {2022},
  doi     = {10.1038/s41598-022-23054-7}
}

\clearpage


\section*{Supplementary material}

# Broadly Tunable Quantum Enhanced Raman Microscopy for Advancing Bioimaging: supplemental document

## 1. SOURCE OF AMPLITUDE-SQUEEZED LIGHT

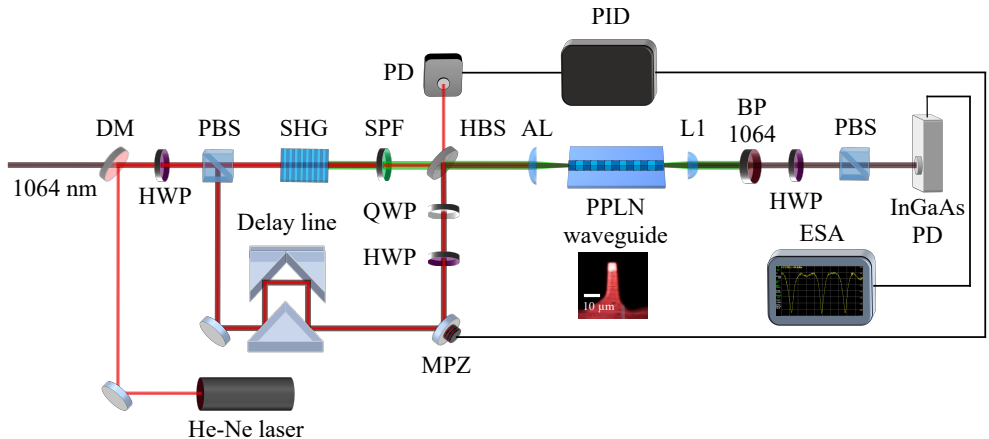

**Fig. S1.** Source of amplitude-squeezed light (Squeezer). DM, dichroic mirror; SHG, second harmonic generator; MPZ, mirror with piezo actuator; QWP, quarter-wave plate; HWP, half-wave plate; PBS, polarizing beam splitter; HBS, harmonic beam splitter, AL, achromatic lens, L1, aspheric lens; BP 1064, 1064 nm band-pass filter; SPF, short-pass filter; InGaAs PD, home-made resonant photodetector; PD, photodetector; ESA, electronic spectrum analyzer.

In our experiment, amplitude squeezing is achieved via traveling-wave optical parametric amplifier (OPA) in a second-order nonlinear waveguide [1, 2]. Figure S1 presents a schematic of the amplitude-squeezed light source. The Stokes beam at 1064 nm from the picoEmerald laser is split into two arms within the setup. In the transmitted arm, the OPA pump is obtained through second-harmonic generation (SHG) using a 7 mm periodically poled lithium niobate (PPLN) crystal with type-0 quasi-phase matching. This pump beam is subsequently coupled into a PPLN waveguide (NTT Innovative Devices, WH-0532-000-A-C-C-TEC), which has a rectangular cross-section of $6 \times 6.5$ $\mu m^2$, via an achromatic lens (AL, Thorlabs AC060-010-B-ML). Inside the waveguide, a squeezed vacuum (SV) is generated in the $TE_{00}$ mode via type-0 parametric down-conversion. The SV is then displaced by a coherent beam (displacing beam) using an asymmetric beam splitter, thereby forming an amplitude-squeezed state suitable for quantum-enhanced SRS measurements. The displacing beam is prepared in the second arm (reflected from the first polarizing beam splitter (PBS, Thorlabs PBS205)) and coupled into the waveguide with the polarization orthogonal to that of the pump, ensuring it does not interact with the generated SV. A zero-order half-wave plate (HWP, Thorlabs WPH05M-1064) and quarter-wave plate (QWP, Thorlabs WPQ05M-1064) are used to adjust the polarization of the displacing beam and SV. A free-space delay line is also introduced to compensate for temporal mismatch between the SV and the displacing beam. After exiting the waveguide through an aspheric lens (L1, Thorlabs C110TMD-B), the two beams were recombined at the transmitted port of an asymmetric beam splitter implemented with a HWP and a PBS. The resulting displaced SV retains the nonclassical noise properties of the SV while incorporating a fraction of the coherent beam amplitude. By adjusting the polarization of the SV prior to the PBS, the effective transmission-to-reflection ratio can be finely tuned, enabling precise control of the displacement operation.

A He–Ne laser (Thorlabs, HNLS008L-EC, $\lambda$ = 632.8 nm) is employed as a phase reference to stabilize the relative phase between the SV and the displacing beam. The He–Ne beam is spatially

overlapped with the incident Stokes beam, and the resulting interference after the harmonic beam splitter (HBS, Thorlabs HBSY12) is detected with a photodiode (PD, Thorlabs DET10A2). The signal is processed by a home-built PID controller, which actively stabilizes the phase via feedback to a piezo-mounted mirror (MPZ). The PID controller also generates sawtooth modulation pulses, enabling controlled scans of the relative phase between the SV and the displacing beam.

However, due to the strong optical field confined within the PPLN waveguide, the total coupled power is strictly limited by the waveguide's damage threshold. In the final experimental configuration, the coherent beam power at the output of the waveguide was $P_{1064\,\text{nm}} = 42$ mW, while the maximum OPA pump power was $P_{532\,\text{nm}} = 2.5$ mW. The total coupling efficiency into the waveguide was approximately 70% for both beams. The transmission/reflection (T/R) ratio of the asymmetric beam splitter was set to 91/9 for SV. As a result, the achived displacement of SV by coherent beam was 3.75 mW.

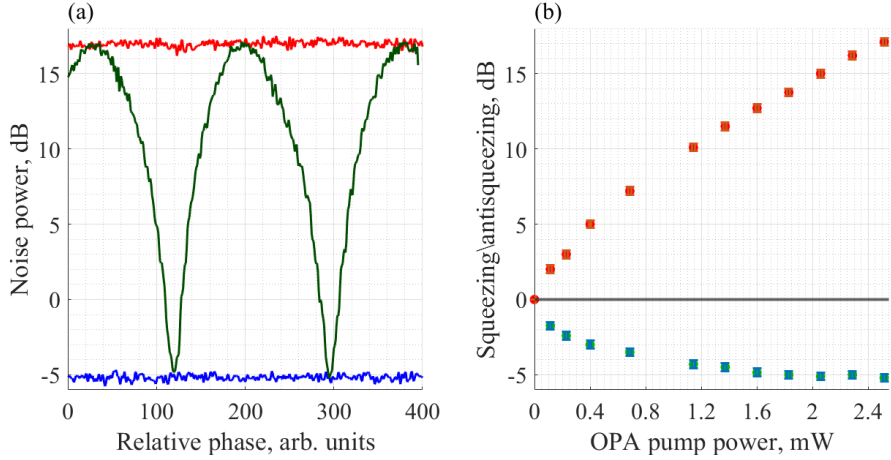

**Fig. S2.** (a) Amplitude noise of the squeezer output vs the seed beam phase at 2.5 mW OPA pump power. (b) Maximal amplitude squeezing and anti-squeezing as functions of the OPA pump power.

Figure S2(a) presents the amplitude noise of the prepared state as a function of the phase of the displacing beam, for the OPA pump power 2.5 mW. At certain phase values, the amplitude noise is $-5.2 \pm 0.2$ dB below the shot-noise level. Panel (b) of the same figure shows the dependence of the measured maximal (antisqueezed) and minimal (squeezed) noise on the OPA pump power. The measurement was conducted using a compact electronic spectrum analyzer (ESA, TinySA Ultra) in combination with a custom-built photodetector (InGaAs PD). The ESA was configured for zero-span acquisition at 19.3 MHz, with a resolution bandwidth (RBW) of 30 kHz, a video bandwidth (VBW) of 300 Hz, and averaging over 250 acquisitions. After the optimal squeezing level is achieved at the OPA pump power 2.5 mW, the noise reduction saturates, primarily due to optical losses in the experimental setup. The total detection efficiency without the microscopy system was estimated as $\eta = \eta_{\text{opt}} \cdot \eta_{\text{sp}} \cdot \eta_{\text{det}} \cdot \eta_{\text{dis}} \approx 0.73$, where $\eta_{\text{opt}} \approx 0.89$ is the total optical transmission of the setup, $\eta_{\text{sp}} = 0.93$ is the spatial overlap between the displacing beam and the SV, $\eta_{\text{det}} = 0.97$ is quantum efficiency of the InGaAs photodiode detector, $\eta_{\text{dis}} = 0.91$ is the transmission of the SV through the PBS used for the displacement.

## 2. QUANTUM ENHANCEMENT OF SRS SIGNAL

To demonstrate the enhancement of stimulated Raman scattering (SRS) sensitivity using amplitude-squeezed light instead of a classical (coherent) Stokes beam, we measured the SRS signal from a 1 mm thick polystyrene layer at the 3050 cm$^{-1}$ vibrational mode. The experimental setup is shown in Figure S3(a). The amplitude-modulated Raman pump at 803.5 nm is combined with the Stokes beam on a dichroic mirror (DM, Thorlabs DMLP1000) after the squeezer module and then focused onto the sample using an objective lens (MO 10×, Thorlabs LMH-10X-1064). After interaction with the sample, the beams are collimated using an achromatic lens (AL, Thorlabs AC050-008-B-ML). Prior to detection, the Raman pump beam is removed with a long-pass filter (LPF , Thorlabs FELH1000), and the Stokes beam is detected using an InGaAs PD photodetector

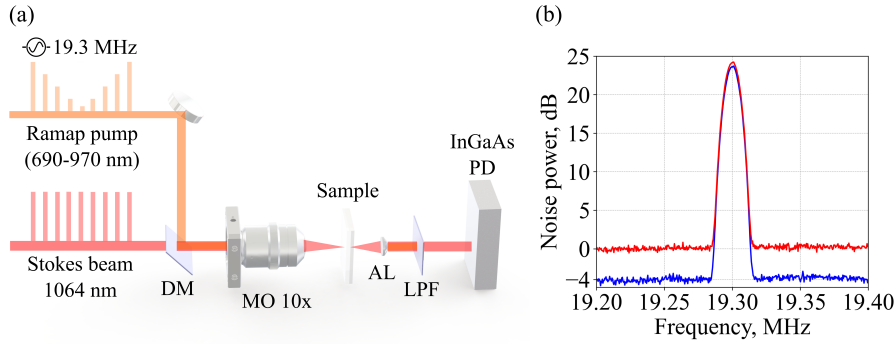

**Fig. S3.** (a) Experimental setup for quantum-enhanced SRS microscopy. DM, dichroic mirror; AL, achromatic lens; LPF, long-pass filter; InGaAs PD, home-made resonant photodetector. (b) SRS signal from the 3050 cm$^{-1}$ vibrational mode of polystyrene measured with coherent (red) and squeezed (blue) Stokes beams.

connected to ESA. The ESA is configured in a 200 kHz span mode with a center frequency of 19.3 MHz, RBW of 30 kHz, VBW of 300 Hz, and averaging over 250 acquisitions. The average optical power of the Raman pump in the sample is 40 mW, while the Stokes beam power is 2.3 mW. Figure S3(b) shows the results of this measurement: clearly, the use of an amplitude-squeezed Stokes beam significantly (by 4 $\pm$ 0.25 dB) reduces the noise and hence improves the sensitivity of the SRS system. To the best of our knowledge, this represent some of the best results reported to date for quantum-enhanced SRS measurements on polystyrene [3].

### REFERENCES

1. T. Kashiwazaki, T. Yamashima, K. Enbutsu, *et al.*, "Over-8-db squeezed light generation by a broadband waveguide optical parametric amplifier toward fault-tolerant ultra-fast quantum computers," Appl. Phys. Lett. **122**, 234003 (2023).
2. Y. Eto, A. Koshio, A. Ohshiro, *et al.*, "Efficient homodyne measurement of picosecond squeezed pulses with pulse shaping technique," Opt. Lett. **36**, 4653–4655 (2011).
3. R. B. de Andrade, H. Kerdoncuff, K. Berg-Sørensen, *et al.*, "Quantum-enhanced continuous-wave stimulated raman scattering spectroscopy," Optica **7**, 470–475 (2020).


\end{document}